\documentstyle[psfig]{mn}

\newcommand{\ltapprox}{\raisebox{-0.5ex}{$\,\stackrel{<}{\scriptstyle
\sim}\,$}}
\newcommand{\gtapprox}{\raisebox{-0.5ex}{$\,\stackrel{>}{\scriptstyle
\sim}\,$}}

\title[Identification of a Gravitationally Lensed z=2.515 Star-Forming Galaxy]
	{Identification of a Gravitationally Lensed z=2.515 Star-Forming Galaxy
	\thanks{Based on observations obtained on the William Herschel
	Telescope at the Observatorio del Roque de los Muchachos, La Palma.}}

\author[T.M.D. Ebbels et. al.]
       {T.M.D. Ebbels$^{1}$, J.-F. Le Borgne$^{2}$, R. Pell\'o$^{2}$,
	R.S. Ellis$^{1}$,\cr J.-P. Kneib$^{1}$,
	I. Smail$^{3}$ \& B. Sanahuja$^{4}$\\
$^{1}$Institute of Astronomy, Madingley Road, Cambridge CB3 0HA, U.K.\\
$^{2}$Observatoire Midi-Pyr\'en\'ees, 14 Av. E.Belin, 31400 Toulouse, France.\\
$^{3}$The Observatories of the Carnegie Institution of Washington, 813 Santa
Barbara St., Pasadena, CA 91101-1292.\\
$^{4}$Departament d'Astronomia i Meteorologia, Universitat de Barcelona,
Diagonal 648, 08028 Barcelona, Spain.\\}

\date{ Accepted 1996
      Received 1996 ;
      in original form }

\begin{document}

\maketitle

\begin{abstract}
We discuss the optical spectrum of a multiply-imaged arc resolved by HST
in the $z$=0.175 cluster A2218. The spectrum, obtained with LDSS-2 on
the 4.2m William Herschel telescope, reveals the source to be a
galaxy at a redshift $z$=2.515 in excellent agreement with the
value predicted by Kneib et al. (1996) on the basis of their inversion of a
highly-constrained mass model for the lensing cluster. The source is
extremely blue in its optical-infrared colours, consistent with active
star formation, and the spectrum reveals absorption lines characteristic
of a young stellar population. Of particular significance is the absence
of Lyman-$\alpha$ emission but the presence of a broad
Lyman-$\alpha$ absorption. The spectrum is similar to that of other,
much fainter, galaxies found at high redshift by various techniques and
illustrates the important role that lensing can play in detailed studies of the
properties of distant galaxies.

\end{abstract}


\section{Introduction}

Considerable progress has been made recently from systematic redshift
surveys of faint galaxies in understanding the evolutionary properties
of galaxies out to redshifts z$\simeq$1. (Colless 1995, Lilly et al.,
1996, Ellis et al.,1996). Beyond this strategic barrier, a variety of
new techniques promise to extend our understanding to higher
redshift. Cowie et al. (1995) have extended the direct spectroscopic
approach for $K$-limited samples to $z\simeq$1.6 and shown a number
have sufficiently strong star formation to account for a high fraction
of the present-day stellar density. A drawback of this approach, as a
way of finding representative distant galaxies is that only a minority
of a magnitude-limited sample, even at the faint limits now reachable
with 10-m class telescopes, lie beyond $z\simeq$1. Presumably even
fainter surveys would be required to uncover significant numbers of
$z>$2 galaxies. Although systematic, this method may not be the most
efficient way to locate high redshift galaxies.

In contrast, Steidel et al. (1996) have shown that very high redshift
sources can be effectively isolated from multiband photometry using
the the Lyman limit as a discriminator. However, the Lyman limit
($\lambda_{rest}$=912 \AA\ ) only becomes amenable to optical
detection at redshifts $z\gtapprox$2.8 and thus, although undoubtedly
a very powerful technique, any comparison of such a distant sample
with one similarly selected at a lower redshift will remain an
observational challenge for the UV capabilities of HST (e.g. Conti et
al 1995). Finally, one must consider the searches for damped Lyman-$\alpha$
absorbers (e.g. Wolfe et al. 1995.) The main disadvantage here is
contamination of the spectrum by the QSO light which inevitably
hampers detailed investigation of the absorbing galaxy. Secondly, one
has to contend with an uneven selection function in redshift produced
by the varying detectability of the absorption.

In this article we illustrate the important role that gravitational
lensing can play in selecting distant galaxies over a large redshift
range, spanning those covered by the methods discussed above.  The
principal advantage of lensing in this context is, of course, the
boosting of the apparent magnitude of the faint galaxy. Early work in
this direction was begun by Smail et al. (1993) who discussed the
photometric properties of giant arcs of known redshift in the context
of various evolutionary models. The unlensed magnitudes of the arcs in
their sample are estimated to be fainter than the galaxies studied in
the deepest current conventional 10m samples (e.g. Cowie et al.  1995)
. Although it is well known that lensing conserves surface brightness,
parts of the source smaller than the seeing disk do undergo a real
surface brightness gain (see Ebbels et al. 1996b). Although this may
not be as large as the formal lensing magnification, the spectroscopic
gain in depth can be significant. Additionally, the spatial magnification
enables distant sources to be resolved into their individual
components using HST (Smail et al. 1996).

The data from HST has enabled the construction of very precise mass
models for selected lensing clusters. A good example is the recent
analysis of Abell 2218 ($z=0.175$) where the ground-based mass model
of Kneib et al (1995) was confirmed in exquisite detail from as many
as 7 multiply-imaged sources resolved by HST in a 3-orbit exposure
(Kneib et. al. 1996, KESCS). The highly-constrained mass model allowed
KESCS to invert the lensing equations and predict redshifts for
individual arclets readily recognised in the HST images.  A number of
the arclets were predicted to have redshifts $z>1$ and, as a way of
isolating high redshift sources, the technique will become more
effective as higher redshift lenses are employed.  As the inversion is
a purely geometric technique, selection effects are less troublesome
than with the alternative approaches. Some of the difficulties
inherent in the optical detection of lensed galaxies are discussed in
KESCS.  Clearly it is important to verify the lensing inversion, where
possible, and in this article we illustrate the potential of the
method via the spectrum of a spectacular multiply-imaged arc in Abell
2218 (\#384 in the numbering scheme of Le Borgne et al, 1992.) The
large magnification (almost 3 magnitudes for this image) allows us to
obtain the spectrum of a normal galaxy at high redshift. The redshift
obtained, $z=2.515$, is in very good agreement with that predicted by
the lensing model of KESCS.

A plan of the paper follows. In \S2 we present the spectroscopic
observations that lead to the redshift determination and discuss
its implications with respect to the inversion method discussed by KESCS.
In \S3 we discuss the broad-band photometric properties
of the distant galaxy and estimate the ongoing star formation rate
and rest-frame luminosities. In \S4 we return to the spectrum and
interpret the various features in the context of the photometric
data and UV spectra of nearby galaxies and recent spectra of
high redshift sources. \S5 summarises the main conclusions of the paper.

\begin{figure}
\psfig{file=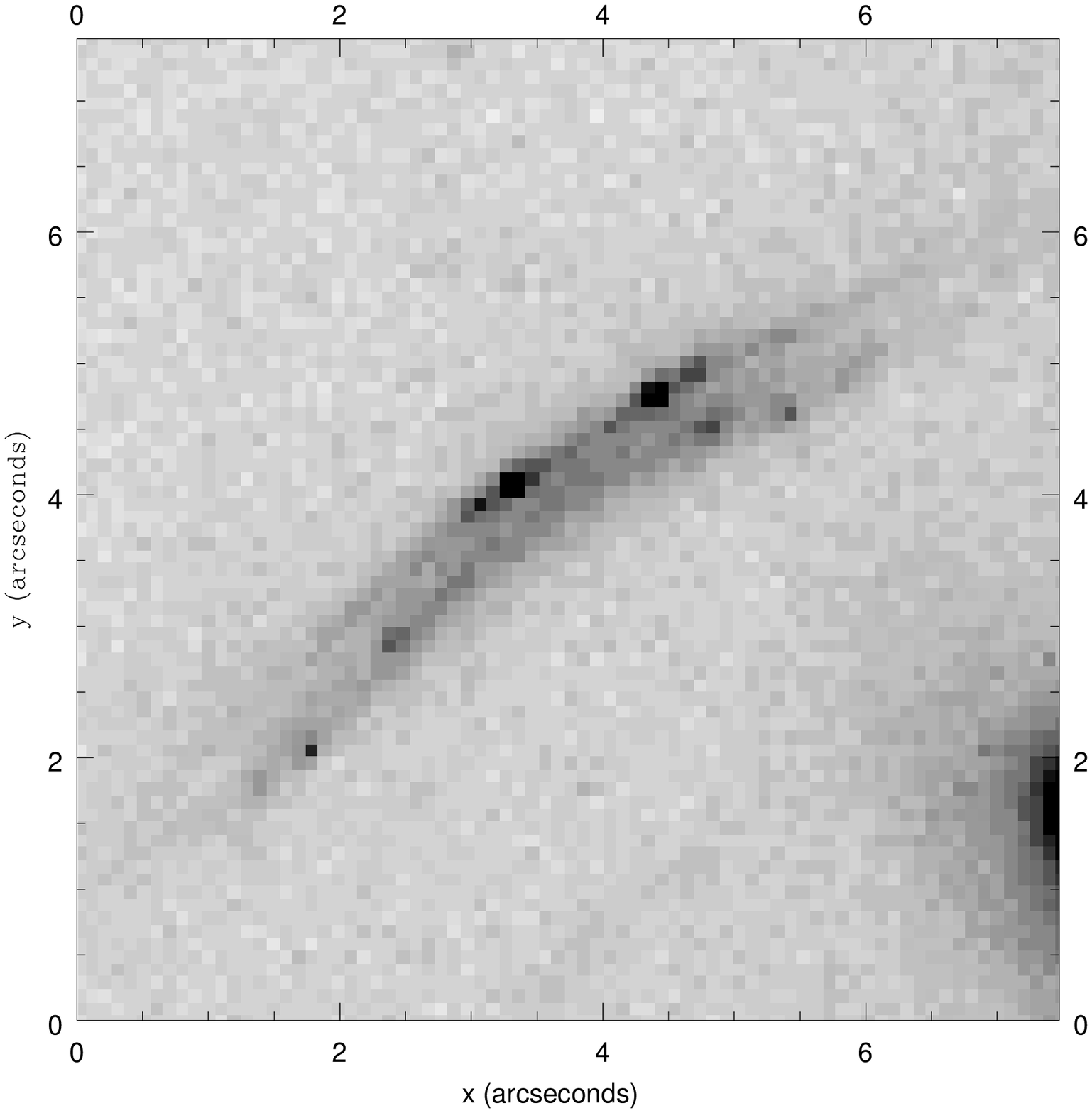,width=0.5\textwidth,angle=0,clip=}
\psfig{file=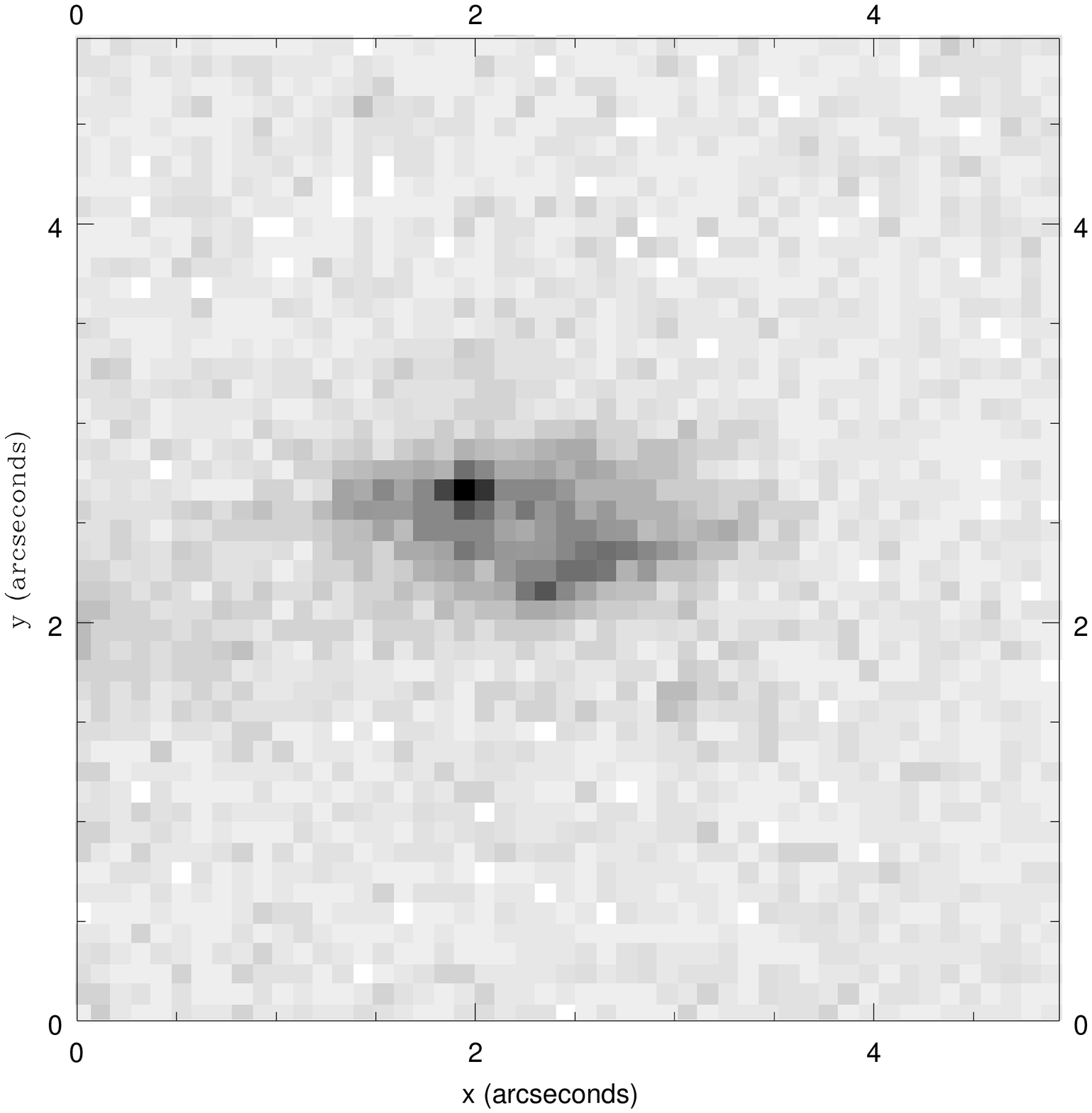,width=0.5\textwidth,angle=0,clip=}
\caption{({\it top}) WFPC-2 F702W image of arc \#384 (after KESCS),
({\it bottom})  WFPC-2 F702W image of arc \#468 considered by KESCS
to be a the counter-image of \#384.}
\label{fig-384}
\end{figure}

\section{Spectroscopic Observations and Redshift Determination}

As part of a major effort to verify the lensing inversion method
for Abell 2218 discussed by KESCS, we have secured
spectra for a large sample of faint arclets using the LDSS-2 multi-slit
spectrograph at the William Hershell Telescope (WHT). A more complete
description of these observations will be presented elsewhere (Ebbels
et al., 1996b). The arclets were selected to be those with
lensed magnitudes $B<$25 for which the inversion gave estimated redshifts
$z_{lensing}>$0.4. Although a colour selection was imposed to increase
the chance of detecting [O II] emission in the spectral window of
LDSS-2, this is immaterial in the present discussion.

The observations were conducted on the nights of May 28 to June 1 1995.
LDSS-2 was used in a conventional multi-slit mode with masks constructed
with straight slits of width 1.5 arcsec configured in two orientations.
The TEK-1 CCD was used (0.59 arcsec pixel$^{-1}$) and in the case of this
arc, a total exposure of 29.9ksec was obtained with the medium blue
grism, (5.3$\AA$ pixel$^{-1}$, $R\simeq 12\AA$) in
conditions with seeing $\simeq$1-2 arcsec FWHM. This grism allows us to
survey the range where Lyman-$\alpha$ was predicted to lie for arc \#384 by
Pell\'o et al. (1992) and Kneib et al. (1995). (3700-5300 \AA\ . This
range was not obtained in Pell\'o et al.'s FOS spectrum.)

Standard data reduction was performed with the LEXT package (Colless et al.
1990, Allington-Smith et al. 1994). After debiasing, the data were
median combined, allowing for flexure effects of $\sim$1 pixel
which affect the long exposures of a single field. Dispersed tungsten
flat fields were used to correct for variations in transmission along the slit,
while wavelength calibration was achieved using a CuAr arc. In extracting the
faint spectra, the light profile along the slit was used to
fit the remaining background variations and isolate the object. Residual
column to column deviations from this fit were removed linearly
and the object rows summed with a gaussian weighting to form an
optimally-extracted spectrum. Finally the spectra were approximately flux
callibrated using the standard LDSS-2 throughput curves.

The arc in question, \#384 (Figure \ref{fig-384}(top)) in the photometric
catalogue of Le Borgne et al. (1992), is one of the most spectacular in
appearance in the HST image presented by KESCS. The mirror
symmetry apparent in the WFPC-2 data verifies the earlier suggestion
(Kneib et al. 1995) that the arc represents the merging of two images
across a critical line whose location can now be precisely determined.
The mass model discussed by KESCS, which is highly constrained by
6 further multiply-imaged features, strengthens the prediction (Kneib
et al. 1995) of a counter-image to \#384 on
the other side of the central cD (object \#468, Figure \ref{fig-384}(bottom))

The location and orientations of the two images allowed KESCS to
predict the inverted redshift and magnification of the source to
reasonably high precision. Indeed, the redshift of \#384 represents a crucial
constraint because of the arc's proximity to the critical line.
On the basis of the
mass model developed, the source is predicted to lie within the range
$z_{lensing}=2.8\pm0.3$, the uncertainty arising from the detail of the mass
model, once constraints of the the other multiply imaged systems in
the cluster are considered. This prediction agrees well with that made from
broad band photometry and 5300-9000 $\AA$ low resolution spectroscopy, of
2.6$\ltapprox z \ltapprox$3.6 by Pell\'o et al. (1992).

The new redshift, when included in the model, does little to change
the predicted magnification of each image (2.9$\pm$0.3 in the case of
the of the combined \#384 images and 1.5$\pm$0.1 in the case of
\#468. The larger error for \#384 results from its proximity to the
critical line, and thus increased sensitivity to its exact position
and the slope of the mass profile (see KESCS). In calculating the
properties of the unlensed source, we use the more reliable
magnification of \#468.)  This leads to an {\it unlensed} $R$
magnitude for \#384/\#468 of 24.1$\pm$0.2.  As KESCS point out, \#384
comprises two images of a source which itself appears to consist of
two parallel components.  \#468 has the same distinctive parallel
morphology, albeit less distorted by the gravitational shear. Since
the source lies so close to a caustic in the source plane, any bright
parts of it lying outside this caustic would be imaged only in
\#468. The absence of any such extra bright knots in \#468 leads us to
conclude that all high surface brightness parts of the source lie
inside the caustic line.

\begin{figure}
\psfig{file=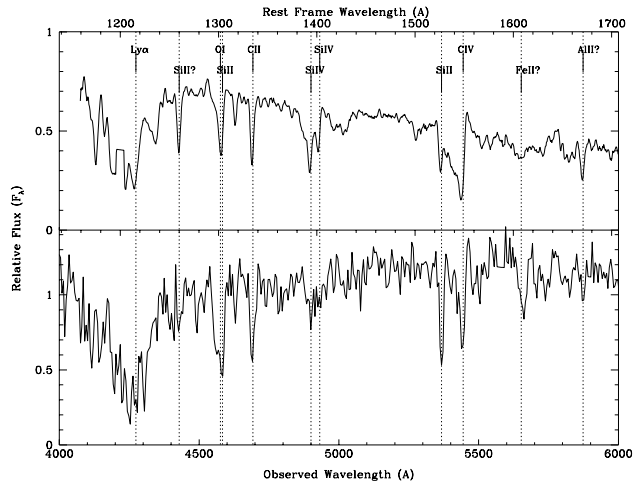,width=0.5\textwidth,angle=270}
\caption{({\it bottom}) LDSS-2 spectrum of the arc \#384. Both strong
and more tentative identifications of features indicating a redshift of
$z = 2.515\pm0.001$ are marked. Sky
residuals near 5577A and 5892A have been truncated for clarity.
({\it top}) For comparison, the HST spectrum of a star forming
knot in NGC 1741
smoothed to the effective resolution of LDSS-2 (Conti et al. 1996).
Many of the same features as in arc \#384 are visable here,
including metal lines and a broad Lyman $\alpha$ which does not
reach the zero flux level. The truncated emission at $\simeq$1200 $\AA$ is
geocoronal Lyman-$\alpha$.}
\label{fig-384spec}
\end{figure}

\begin{figure}
\psfig{file=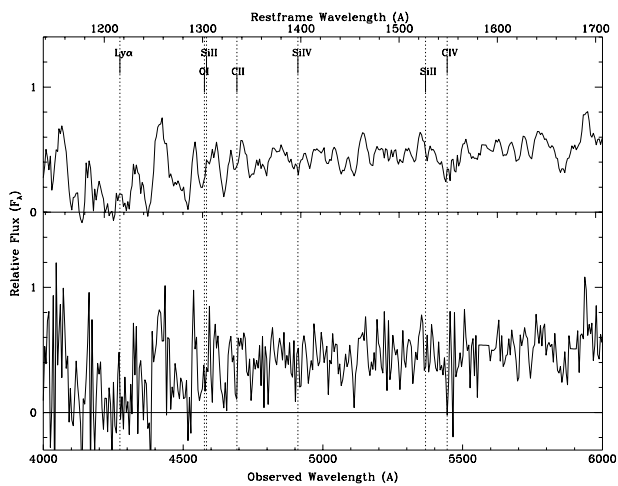,width=0.5\textwidth,angle=270}
\caption{({\it bottom}) LDSS-2 spectrum of the arc \#468 showing that it is
compatible with
that of \#384. ({\it top}) \#468 spectrum smoothed with a boxcar of
width 5 pixels. Positions of stronger lines appearing on the \#384 spectrum are
marked. Solid horizontal lines mark the zero levels. }
\label{fig-468spec}
\end{figure}

The LDSS-2 spectrum for \#384 is shown in Figure \ref{fig-384spec} and
has a very satisfactory signal/noise. Although the integrated
photometry indicates $R_{tot}$=21.2, the mean surface brightness is
only $\mu_R$=23.4 ($\simeq$7 per cent of the La Palma night sky). Thus
the observation represents a considerable challenge. The spectrum
shows a number of strong ultraviolet features consistent with a
redshift $z$=2.515, close to the predicted value. Most noticeable is
the prominent feature at $\simeq$4250 \AA\ which arises from strong
Lyman-$\alpha$ absorption. No emission lines are seen in the optical
spectrum. With our new data, we find that multiple image hypothesis is
reinforced by comparing the spectra of all three images. The two
halves of \#384 show no difference in spectral features, while the
spectrum of \#468 (figure \ref{fig-468spec}), although noisy, is
certainly compatible with that of \#384.  We return to a detailed
discussion of the spectral features in the next section.

The verification of the inverted redshift, to within the estimated
precision, illustrates the power of the lensing technique for
isolating high redshift galaxies as well as for determining the
statistical redshift distribution of very faint sources. It also shows
that the mass model for the cluster core is very well determined, and
in fact, further constrains it. The implications of the comparison of
the lensing predicted redshifts and spectroscopically determined ones
will be presented in Ebbels et al. (1996b), where the spectroscopic
results from all the arclets secured in the WHT run are collated and
analysed.

\section{Photometric Properties}

\begin{table*}
\centering
\caption{Optical and Infra-red photometry of \#384 and \#468. Values for
$\mu_R$ \& $R$ are taken from KESCS, those
for $J$ \& $K'$ from Kneib
et al. (1995) and those for $grz$ from Le Borgne et al. (1992),
where the likely errors are
$\pm$0.1, $\pm$0.4 and $\pm$0.3 respectively. Values for $UBVi$ are
taken from new photometry obtained at the 5m Palomar telescope (TEK CCD,
seeing 1-1.5 arcsec, exposures of 2.5,16.5,5.8,21.7 ksec respectively).
Estimated errors are $\pm$0.1 in BVi and $\pm$0.3 / $\pm$0.6 in U for \#384
/ \#468 respectively.
Also listed is the relative magnification, $\delta_m=(m_{384}-m_{468})$ of the
two images. The unlensed magnitude
of the source may be obtained by adding the magnification $M$ to the values in
the table, where $M_{384}$=2.9$\pm$0.3 and $M_{468}$=1.5$\pm$0.1}
\label{tab-phot}

\begin{tabular}{ccccccccccccc}
\noalign{\smallskip}
\hline
\noalign{\smallskip}
\# & $\mu_R$ & U & B & g & V & R & r & i & z & J & K' \\
\noalign{\smallskip}
\hline
\noalign{\smallskip}
384 & 23.4 & 22.2 & 22.2 & 22.1 & 21.4 & 21.2 & 21.6 & 20.6 & 21.5 & 20.0 & 17.
9 \\
468 & 23.5 & 23.8 & 23.8 & 23.6 & 23.1 & 22.6 & 23.1 & 22.3 & 22.9 & 20.9 & 19.
1
 \\
$\delta_m$ & & 1.6 & 1.6 & 1.5 & 1.7 & 1.4 & 1.5 & 1.7 & 1.4 & 0.9 & 1.2 \\
\noalign{\smallskip}
\hline
\noalign{\smallskip}
\end{tabular}
\end{table*}

Although there are now several examples of non-active galaxies with
redshifts above 2, most can only be seen by virtue of their abnormal
luminosities. The significant boosting of the source responsible for
the arcs \#384+\#468, together with the serendipitous way in which
background galaxies are lensed makes lensing a useful probe of normal galaxies
at hitherto unexplored epochs, particularly if, as seems clear,
further examples can be found in this way. We now illustrate, using
this source as an example, what can be learnt from the photometric
and spectroscopic properties of lensed high redshift galaxies.

Multiband photometry from U to K' is presented for both \#384 and
\#468 in Table 1 where the revised magnification is also given.  The
most reliable method to estimate the source luminosity is to use the
infrared photometry to estimate the rest-frame optical
luminosity. This is most suitably done with the $J$ or $K'$ photometry
which can yield the rest-frame $U$ or $R$ luminosity to reasonable
precision. As $(J-K')_{obs}$ maps closely onto $(U-R)_{rest}$, it can
be seen from table \ref{tab-phot} that the galaxy is intrinsically
very blue, corresponding to a flat-spectrum source. To obtain the
intrinsic $b_{Jrest}$ luminosity of the source, a colour correction
from $U_{rest}$ or $R_{rest}$ to $b_{Jrest}$ is required and this
yields $M_{b_J}$=-22.2 (-23.0) $\pm$0.5 corresponding to a galaxy
$\simeq$8-20 (16-40) $h_{50}^{-2}$ $L_{b_J}^{\ast}$ for $q_0$=0.5
(0.1). (Loveday et al, 1992.) However, the blue magnitude is much
influenced by star formation and, given its blue colours
($(U-R)_{rest}\simeq(J-K')_{obs}$=2.1), it is possible that the rest
frame K luminosity of our galaxy could be closer to $L^{\ast}$ than
suggested by the above $M_{b_J}$ value.

\begin{figure}
\psfig{file=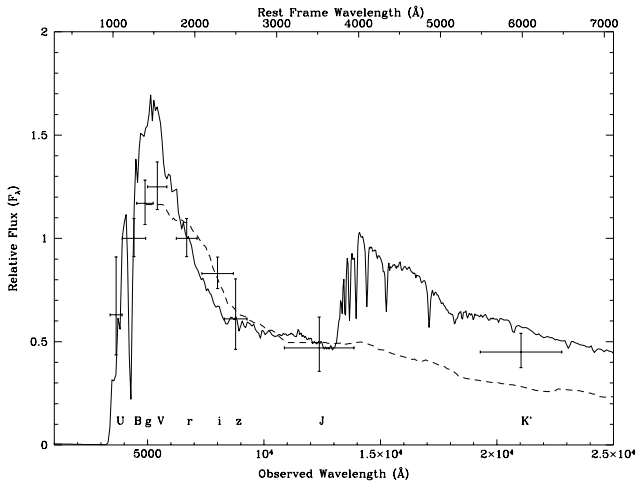,width=0.5\textwidth,angle=270}
\caption{The spectral energy distribution of the $z$=2.515
galaxy as determined from available broad-band photometry. The solid
line represents the spectral energy distribution of a model
of the burst type with no reddening and age of 0.286Gyr. The dotted
line shows the smoothed spectral energy distribution of NGC 4449, a well
studied local example of an actively star forming galaxy.}
\label{fig-384sed}
\end{figure}

The spectral energy distribution (SED) delineated by the broad-band
photometry is plotted in Figure \ref{fig-384sed} where we compare it
to a synthetic spectrum produced by Bruzual \& Charlot's (1993) code
and to the smoothed SED of a nearby star forming galaxy, NGC 4449
(Ellis 1984).  These both show the SED to be consistant with that of
an actively star forming galaxy. Although the best model fit obtained
was that of a burst (of age $\simeq 3 \times 10^8yr$, no reddening),
other models (including those with constant star formation) can also
reproduce the gross features of the photometry, particularly given the
freedom to invoke reddening, which has sizable effects at UV
wavelengths (the exact nature of which is little known at these
redshifts).  At this redshift, the Lyman limit only just reaches the
edge of the U band. Thus the drop in the photometry, although
influenced by the Lyman break, cannot easily be compared with the
($U_n$-$G$) cut from Steidel et al. (1996).  Following Steidel et al,
the UV flux can be used to derive an approximate ongoing star
formation rate. In the case of a Salpeter initial mass function,
(slope $\alpha$=2.35, with $1M_{\odot} \leq M \leq 80M_{\odot}$), this
amounts to $\simeq$7-11 (15-24)$h_{50}^{-2}$ $M_{\odot}yr^{-1}$ for
$q_0$=0.5 (0.1), which is similar to the range seen in the Steidel
sample and comparable to that in NGC 4449 when allowance for
luminosity differences is made ($\simeq 8
M_{\odot}yr^{-1}$. (Thronson, 1987))

As only the HST has the spatial resolution to adequately
detect internal structures, and the WFPC-2 data is only available in
a single band (F702W), no colour variations along arc \#384 can
be seen. However, the presence of `knots' in Fig. \ref{fig-384}
is suggestive of luminous HII regions consistent with a galaxy
undergoing strong star formation.

\section{Spectroscopic Properties}

The conclusions based on multiband photometry are also supported
by analysis of the spectrum of \#384 (Figure \ref{fig-384spec}).
The most striking features are the broad Lyman-$\alpha$ absorption and
the presence of several ultraviolet metal absorption lines. The latter
features are typical of UV lines seen with IUE in the spectra of hot stars
and nearby star forming galaxies such as those catalogued by Kinney et al
(1993). As an illustration, Figure \ref{fig-384spec} also shows the HST
spectrum of a starburst knot in NGC 1741, a Wolf-Rayet galaxy at a
distance of 77$h_{50}^{-1}$Mpc. (Conti et. al. 1996). The spectral lines
seen are similar to those in our source, while the luminosity of this knot
at 1500 $\AA$ is a factor $\simeq$10 fainter (as in the Steidel et
al. (1996) sample).

Higher resolution data are required for a detailed analysis of the spectrum
(e.g. definative separation of stellar / interstellar absorption) but
our data are sufficient to reveal several interesting aspects of
this faint object. Using the redshift determined
from the metal lines, it can be seen that the Lyman-$\alpha$ line
is somewhat assymmetric (by $\simeq$200$\AA$) on its blueward side.
This may possibly be explained by a blend with metal lines such as
SiIII 1207$\AA$. There is also some evidence of further absorption at shorter
wavelengths which could be Lyman-$\alpha$ absorption from foreground material.
The Lyman-$\alpha$ absorption does not seem to reach
a zero flux level although the signal to noise in the blue remains poor.
Similar cases are seen in Kinney et al's sample, in the HST spectra of
nearby galaxies discussed by Leitherer et al. (1996) \& Conti
et al. (1996), and in the high redshift
spectra of Steidel et al. (1996) and Yee et. al. (1996). The claimed
luminosity of the latter example is a factor of $\simeq$130 (at 1500$\AA$)
brighter than our source, strongly suggesting gravitational amplification by
the
cluster through which it is observed. This large factor allows the aquisition
of a much higher signal to noise spectrum than the one presented here.

The strong stellar and interstellar lines seen by the above authors also
appear strong in our spectrum with a few exceptions, the most notable of
these being the weakness of SiII 1260$\AA$ c.f. SiII 1527$\AA$. Although
better signal to noise data are required, this ratio is the reverse of
that seen in stellar spectra and raises the question of whether the 1527$\AA$
line is contaminated e.g. by a second component possibly associated with
the double nature of the source mentioned in \S2. From analyses of IUE
stellar spectra and interstellar absorption of Magellanic Cloud stars
(e.g. Savage \& de Boer, 1981) we conclude that most of the low ionisation
lines (e.g. SiII, OI, FeII, CII) are due to interstellar
absorption while those of higher ionisation (e.g. CIV, SiIV) are mainly
stellar, from late O to early B stars. In stars later than about B2, the
doublet of 1526.7\AA\,1533.4\AA\ becomes important and would have been resolved
in our spectrum (c.f. Rountree \& Donneborn, (1991)), while in the interstellar
medium, only the line at 1526.7\AA\ is seen (as we observe).
The absence of Lyman-$\alpha$ emission in
a source with demonstrably active star formation confirms earlier
suspicions that such emission can be attenuated by a number of processes
(Charlot \& Fall 1991,1993). It is important to distinguish the configuration
of
our system with
the QSO absorption scenario. In the QSO case, Lyman-$\alpha$ photons
are scattered out of the beam by the absorber. In our case, the sources
are embedded within the scattering medium and (as highlighted by
Charlot \& Fall), the extent of the attenuation depends on the relative
abundances and geometry of the mixture of stars, gas and dust.

\section{Conclusions}

We have secured the redshift, $z$=2.515, of a multiply-imaged arc in
the rich cluster Abell 2218 which we believe to be the first
comfirmation of a redshift predicted by a cluster lensing model. The
result strongly vindicates the precision of the lensing inversion
technique developed by KESCS and further illustrates how gravitational
lensing through well-constrained clusters can locate samples of high
redshift galaxies. The efficiency of this selection technique may be
relatively low, given the surface density of high redshift galaxies
found by Steidel et al. and the cluster cross section, coupled with
the need for high resolution imaging. However, this method is less
susceptible to luminosity biases and redshift selection effects than
other techniques.

The source responsible for the lensed images appears to be a blue 8-20
$L_{bJ}^{\ast}$
galaxy whose on-going star formation rate of 7-11 $M_{\odot}$yr$^{-1}$
is similar to that of similar sources found at higher redshift by
Steidel and collaborators using the Lyman limit cutoff as a high $z$
locator. The spectrum reveals UV absorption lines which are
consistent with the photometric properties. In particular, the Lyman
$\alpha$ line is particularly broad suggesting neutral gas is
abundant within the galaxy. In conclusion, we believe that the spectral and
photometric properties of this source are consistant with those expected
of the objects responsible for high redshift absorption along lines of sight
to distant quasars.

\section*{Acknowledgments}

We thank Claus Leitherer, Howard Yee and Chuck Steidel for sharing
their results with us prior to their publication.
We acknowledge valuable discussions on the interpretation
of ultraviolet spectra and photometry with Bob Carswell, Max Pettini
and Alfonso Arag\'on-Salamanca. We especially note the help Karl Glazebrook
provided as a regular user of LDSS-2, which
operates successfully
thanks to the dedicated efforts of La Palma
support staff and, in particular, Mike Breare. We are grateful to G.
Bruzual for allowing the use of his code for the spectral evolution of
galaxies. We also acknowledge financial support from the Particle Physics
and Astronomy Research Council (TMDE \& IS), European Commission (JPK)
and the DGICYT (Ministerio de Eduacion
y Ciencia, Spain) (BS).


\begin{thebibliography}{}

\bibitem[\protect\citename{Allington-Smith et al.}1994]{allingtonsmith94}
Allington-Smith J., Breare M., Ellis R.S., Gellatly D., Glazebrook K.,
Jorden P., Maclean J., Oates P., Shaw G. \& Tanvir N., 1994, PASP, 106, 983.

\bibitem[\protect\citename{Bruzual \& Charlot}1993]{bruzual93} Bruzual A.,
\& Charlot S., 1993, ApJ, 405, 538.

\bibitem[\protect\citename{Charlot \& Fall}1991]{charlot91} Charlot S.,
\& Fall M., 1991, ApJ, 378, 471.

\bibitem[\protect\citename{Charlot \& Fall}1993]{charlot93} Charlot S.,
\& Fall M., 1993, ApJ, 415, 580.

\bibitem[\protect\citename{Colless et al.}1990]{colless90} Colless M.,
Ellis R.S., Taylor K., \& Hook R.N., 1990, MNRAS, 253, 686.

\bibitem[\protect\citename{Colless}1995]{colless95} Colless M., 1995, in
{\it Wide Field Spectroscopy}, eds. Arag\'on-Salamanca, A. \& Maddox,
S J, World Scientific, p263.

\bibitem[\protect\citename{Conti et al.}1996]{conti96} Conti P.S.,
Leitherer C. \& Vacca W.D., 1996, preprint.

\bibitem[\protect\citename{Cowie et al.}1995]{cowie95} Cowie L.L.,
Hu E.M., \& Songaila A., Nature 377, 603.

\bibitem[\protect\citename{Ebbels et al.}1996b]{ebbels96b} Ebbels
T.M.D., Ellis R.S., Kneib J.-P., Pell\'o, R., \& Le Borgne, J.-F.,
1996b, in preparation.

\bibitem[\protect\citename{Ellis et al.}1994]{ellis94} Ellis R.S., 1984,
in {\it Spectral Evolution of Galaxies}, ed. Gondhalekar, P.,
RAL Publications, p122.

\bibitem[\protect\citename{Ellis et al.}1996]{ellis96} Ellis R.S.,
Colless M., Heyl J.S., Broadhurst T.J., \& Glazebrook K., 1996,
MNRAS, in press

\bibitem[\protect\citename{Kinney et al.}1993]{kinney93} Kinney A.L.,
Bohlin R.C., Calzetti D., Panagia N. \& Wyse R.F.G., 1993, ApJS, 86, 5.

\bibitem[\protect\citename{Kneib et al.}1995]{kneib95} Kneib, J.-P.,
Mellier, Y., Pell\'o, R., Miralda-Escud\'e, J., Le Borgne, J.-F.,
B\"ohringer, H., Picat, J.-P., 1995, A\&A, 303, 27.

\bibitem[\protect\citename{Kneib et al.}1996]{kneib96} Kneib, J.-P.,
Ellis, R.S., Smail, I.R., Couch, W.J., Sharples, R.,
1996, ApJ ~submitted (KESCS).

\bibitem[\protect\citename{Le Borgne et al.}1992]{leborgne92} Le
Borgne, J.-F., Pell\'o, R. \& Sanahuja, B., 1992, A\&AS, ~95, 87.

\bibitem[\protect\citename{Leitherer et al.}1996]{leitherer96}
Leitherer C., Vacca W.D., Conti P.S., Filippenko A.V., Robert C. \&
Sargent W.L.W., 1996, ApJ, submmitted.

\bibitem[\protect\citename{Lilly et al.}1996]{lilly96} Lilly, S.J., Le
Fevre, O., Hammer F.\& Crampton D., ApJ Letters in press.

\bibitem[\protect\citename{Loveday et al.}1992]{loveday92} Loveday J.,
Peterson B.A., Efstathiou G. \& Maddox S.J., 1992, ApJ, 390, 338.

\bibitem[\protect\citename{Pell\'o et al.}1992]{Pello92} Pell\'o R.,
Le Borgne, J.-F., Sanahuja, B., Mathez G., \& Fort B., 1992, A\&A,
266, 6.

\bibitem[\protect\citename{Rountree \& Donneborn}1993]{Rountree93} Rountree J.,
\& Donneborn G., 1991, ApJ, 369,515

\bibitem[\protect\citename{Savage \& de Boer}1981]{savage81} Savage B.D.
\& de Boer, K.S., 1981, ApJ, 243, 460.

\bibitem[\protect\citename{Smail et al.}1993]{smail93} Smail I.,
Ellis R.S., Arag\'on-Salamanca A., Soucail G., Mellier Y., \& Giraud E.,
1993, MNRAS, 263, 628.

\bibitem[\protect\citename{Smail et al.}1996]{smail96} Smail I.,
Dressler A., Kneib J.-P., Ellis R.S., Couch W.J., Sharples R.M.
\& Oemler A., 1996, ApJ., in press.

\bibitem[\protect\citename{Steidel et al.}1996]{steidel96} Steidel C.C.,
Giavalisco M., Pettini M., Dickinson M., \& Adelberger K.L., 1996, preprint.

\bibitem[\protect\citename{Thronson et al.}1987]{thronson87} Thronson H.A.,
Hunter D.A., Telesco C.M., Harper D.A., \& Decher R., 1987, ApJ, 317, 180.

\bibitem[\protect\citename{Wolfe et al.}1995]{wolfe95} Wolfe A.M.,
Lanzetta K.M., Foltz C.B. \& Chaffee F.H., 1995, ApJ, 454, 698.

\bibitem[\protect\citename{Yee et al.}1996]{yee96} Yee H.K.C., Ellingson E.,
Bechtold J. \& Carlberg R.G., 1996, preprint.

\end{thebibliography}
\end{document}